\documentclass[conference,10pt]{IEEEtran}
\IEEEoverridecommandlockouts
\usepackage{cite}
\usepackage{amsmath,amssymb,amsfonts}
\usepackage{algorithm}
\usepackage{algorithmicx}
\usepackage{algpseudocode}
\usepackage{graphicx}
\usepackage{textcomp}
\usepackage{xcolor}
\usepackage{lettrine}
\usepackage{multirow}
\usepackage{caption}
\usepackage{subcaption}
\usepackage{tabularx}
\usepackage{romannum}
\usepackage{array}
\usepackage{cleveref}
\usepackage{url}
\crefrangeformat{figure}{Fig.~#3#1--#4#2}
\newcolumntype{P}[1]{>{\centering\arraybackslash}p{#1}}
\newcolumntype{M}[1]{>{\centering\arraybackslash}m{#1}}
\def\BibTeX{{\rm B\kern-.05em{\sc i\kern-.025em b}\kern-.08em
    T\kern-.1667em\lower.7ex\hbox{E}\kern-.125emX}}
\begin{document}
\setlength{\abovedisplayskip}{3pt}
\setlength{\belowdisplayskip}{3pt}
\setlength{\floatsep}{0 pt}
\setlength{\textfloatsep}{0 pt}

%\title{Performance Analysis of IRS-Assisted SSK with Reflection Phase Modulation Over Rician Fading
\title{On the Performance of IRS-Assisted SSK and RPM over Rician Fading Channels
\vspace{-0.28 em}
}
%title{A Performance Analysis of IRS-Aided SSK and RPM over Rician Fading Environment}
%\title{ An Analysis on the Performance of IRS-enabled SSK with Reflection Phase Modulation Over Rician Channel}
%\title{}
\author{\IEEEauthorblockN{Harsh Raj$^{\dagger}$, Ugrasen Singh$^{\$}$, and B. R. Manoj$^{\dagger}$\\$^{\dagger}$Department of Electronics \& Electrical Engineering, Indian Institute of Technology Guwahati, Assam, India\\ 
$^{\$}$Department of Information and Communications Engineering, Aalto Univerisity, Helsinki, Finland\\Emails: \text{\tt\{h.raj, manojbr\}@iitg.ac.in, ugrasen.singh@aalto.fi}
\vspace{-2.65 em}
}

\thanks{This work was supported in part by SERB Start-Up Research Grant Scheme, Govt. of India under Grant SRG/2022/001214 and in part by Nokia Solutions and Networks Oy.}
}

\maketitle
\begin{abstract}
This paper presents the index modulation, that is, the space-shift keying (SSK) and reflection phase modulation (RPM) schemes for intelligent reflecting surface (IRS)-assisted wireless network. IRS simultaneously reflects the incoming information signal from the base station and explicitly encodes the local information bits in the reflection phase shift of IRS elements. The phase shift of the IRS elements is employed according to local data from the RPM constellation. A joint detection using a maximum-likelihood (ML) decoder is performed for the SSK and RPM symbols over a realistic fading scenario modeled as the Rician fading channel. The pairwise error probability over Rician fading channels is derived and utilized to determine the average bit error rate. In addition, the ergodic capacity of the presented system is derived.  
The derived analytical results are verified and are in exact agreement with Monte-Carlo simulations.
\end{abstract}

\begin{IEEEkeywords}
Average bit error rate, ergodic capacity, IRS, ML-detector, SSK, reflection phase modulation, Rician fading.
\end{IEEEkeywords}

\section{Introduction}
Intelligent reflecting surface (IRS) is an emerging technology for 6G wireless networks to increase the coverage range, link reliability, and spectral efficiency (SE). Compared to a conventional relay network with one or more radio frequency (RF) chains and a constant power supply for re-broadcasting the incoming signals, IRS does not require RF chains and a regular power supply to direct the incoming beam towards the desired receiver \cite{Tang, Renzo, mukherjee_grouping_2023}. These advantages of IRS over traditional relays have motivated researchers to utilize them in active research of wireless communications.

Index modulation (IM) has been introduced in wireless communications to meet the demands of higher SE and energy efficiency (EE) \cite{Mao_novel_index_2019}. The antenna selection-based IM scheme can be implemented in multiple-input multiple-output (MIMO) systems using a few RF chains. The underlying benefits of IM and IRS have led researchers to combine them to make their benefits available in wireless network design. An IRS-assisted IM scheme is implemented in \cite{Basar_index_modulation_2017}, exploiting the transmit antenna (TA) index and the conventional signal constellation. The space-shift keying (SSK) modulation was introduced with IRS in  \cite{Canbilen_Reconfigurable_Intelligent_2020}, where the IRS is considered as a passive relay. On the contrary, an information transmission scheme through IRS is presented in \cite{Basar20,Yuan21} by utilizing the receive antenna indices, where the IRS directs beam of radio waves towards the selected receive antenna. However, designing a highly directive beam toward the selected receiving antennas from the IRS is challenging. Moreover, simultaneous information transmission and beamforming at the IRS is performed in \cite{Yan20}, where IRS elements' ON/OFF states implicitly convey the information. However, in this scheme, the IRS is partly activated, leading to reduced received signal power at the user. To improve the received signal strength, a reflection pattern modulation scheme is proposed in \cite{lin_Reconfigurable_intelligent_2020}, where IRS implicitly encodes its information in the index of sets of ON-state elements of IRS. The above-discussed IM schemes that exploit the ON/OFF states of the IRS elements activate the portion of IRS at a time, which leads to reduced aperture gain provided by the IRS, resulting in reduced received signal strength.

To achieve the full aperture gain inherent in an IRS, we exploit the reflection phase modulation (RPM) scheme, which activates all the IRS elements at the same time and explicitly implants local information in the reflection phase shift of IRS elements while reflecting the incoming signal. Further, the base station (BS) exploits the SSK modulation scheme to transmit the local data. The authors in \cite{singh_RIS_assisted_2022} considered an IRS system with SSK and RPM scheme to analyse the performance over Rayleigh fading channels. Different from  \cite{singh_RIS_assisted_2022}, we assume a realistic fading scenario where both line-of-sight (LoS) and non-LoS (NLoS) paths of radio waves are considered to measure the received signal strength. Since the IRS is more effective in improving the received signal strength in the presence of LoS scenarios, this motivated us to develop an analytical framework to investigate the performance of IRS-assisted SSK with RPM system over Rician fading channels. 

The main contributions of this paper are as follows: (a) A joint decoding of the SSK and the RPM symbols is performed at the receiver using a maximum-likelihood (ML) detector. (b) We have proposed a moment-generating function (MGF)-based framework for deriving pairwise error probability (PEP). The PEP is derived for the antenna index and phase modulation mismatches over the Rician channels. We encounter the distribution of the sum of squares of Rician random variables (RVs) while evaluating the PEP, for which we have derived an approximation to overcome the analytical and computational complexities. (c) The PEP is then utilized to derive the average bit error rate (ABER) of the considered IRS-SSK-RPM system. A closed-form expression is also derived for the ergodic capacity. (d) We have evaluated the derived results numerically and verified them through simulations to analyze the system performance for parameters such as the number of IRS elements, transmit and receive antennas, size of RPM constellation, and distances between IRS and transceivers.  
\vspace{-0.5 em}
\section{System Model}\label{sec:1}
\vspace{-0.5 em}
A generic IRS-assisted MIMO wireless network is considered, where BS has $N_t$ TAs and the user terminal (UT) has $N_r$ receive antennas. BS exploits the SSK modulation scheme to transmit the information to UT, while IRS has a control unit that configures the reflection phase shift of IRS elements according to the local information bits. 
The control unit has built-in sensors that collect data from the environment. These data are mapped into the RPM constellation to convey information by controlling the reflection phase shift of IRS elements. 
Moreover, IRS is situated in the $x\text{-}y$ plane comprising a total of $N=N_x \times N_y$ reflecting elements, with $N_x$ and $N_y$ denoting the number of elements in the $x$ and $y$ axes, respectively. We assume that the direct link between BS and UT is blocked due to the large obstacle. Further, SSK transmission from BS is represented as $\mathbf{x}=\mathbf{i}_t$, where $\mathbf{i}_t$ denotes the $t$-th column of the $N_t\times N_t$ identity matrix. The $M$-ary RPM constellation is given by $\varphi={\{\varphi_1, \dots, \varphi_M\}}$, where $\varphi_m={\frac{2\pi(m-1)}{M}}, \forall $ $m \in [1, M] $, and the reflection phase shift matrix of IRS is denoted by $\mathbf{\Phi}= \text{diag}\left([e^{j\varphi_1}, e^{j\varphi_2},\dots,e^{j\varphi_M}]\right)$. The channel matrices for the BS-to-IRS and IRS-to-UT links are denoted as $\mathbf{H}\in \mathbb{C}^{{N}\times {N_t}}$ and ${\mathbf{G}} \in {\mathbb{C}^{{N}\times{N_r}}}$, respectively.
\vspace{-1.5 em}
\subsection{Channel Model}\label{sec:1a}
\vspace{-0.5 em}
The LoS component typically carries maximum channel power, whereas the NLoS component is a random stochastic process and carries minimum channel power. Consequently, we consider the strong LoS path for the BS-to-IRS channel given as,  
$\mathbf{H}=\sqrt{\nu} \, \mathbf{a}_\mathrm{IRS}(\phi_a,\phi_e)\mathbf{a}_\mathrm{BS}(\phi_d)$, where $\nu=\rho_0d_{t}^{-\eta}$ represents the path loss of the BS-to-IRS link with $d_t$ being the link distance, $\eta$ is the path loss exponent, and $\rho_0$ is the path loss at a reference distance $d_0$. The antenna array response of IRS $\mathbf{a}_\mathrm{IRS}(\phi_a,\phi_e)$ is given by 
\cite{Wang_Channel_Estimation_2019}
\begin{eqnarray}\label{eq:a}
     \mathbf{a}_\mathrm{IRS}(\phi_a,\phi_e)&\!=\left[1,\ldots,e^{-j2\pi \frac{ \kappa }{\lambda}(n_x\sin{\phi_e}\cos{\phi_a}+n_y\cos{\phi_e})},\right. \nonumber\\
    & \!\!\!\!\! \!\!\!\!\! \!\!\!\!\! \left.\ldots,e^{-j2\pi \frac{ \kappa }{\lambda}((N_x-1)\sin{\phi_e}\cos{\phi_a}\!+(N_y-1)\cos{\phi_e})}\!\right]^\text{T}\, ,
\end{eqnarray}
where $\phi_a \in (0,2\pi)$ and $\phi_e \in (0,2\pi)$ are the azimuth and elevation angle of arrival (AoA) of signal to the IRS, respectively, $ 0 \leq n_x\leq N_x-1$, $0 \leq n_y\leq N_y-1$, $\lambda$ is the wavelength, and $\kappa$ is the spacing between reflecting elements. A uniform linear array vector at the BS is given by   
\begin{equation}\label{eq:b} 
    \mathbf{a}_\mathrm{BS}(\phi_d)=\left[1,e^{-2\pi \frac{\delta}{\lambda}\sin{\phi_d}},\ldots,e^{-2\pi \frac{\delta}{\lambda}(N_t-1)\sin{\phi_d}}\right]\, ,
\end{equation}
where $\phi_d \in (0,2\pi)$ is the angle of departure (AoD) from the BS and $\delta$ is the antenna spacing at the BS. We consider the Rician distribution for the IRS and UT channel, expressed as
\vspace{-0.25 em}
\begin{equation}\label{eq:c} 
    \mathbf{G}=\sqrt{\frac{K_r\nu_r}{1+K_r}}\Bar{\mathbf{G}}+\sqrt{\frac{\nu_r}{1+K_r}}\widetilde{\mathbf{G}}\,,
\end{equation}
where $\Bar{\mathbf{G}}$ and $\widetilde{\mathbf{G}} \sim {\cal{CN}}(0, \mathbf{I}_{Nr})$ are the LoS and NLoS components of the channel matrix ${\mathbf{G}}$, respectively,
$K_r$ is the Rician $K$-factor, and $\nu_r=\rho_0{d_r}^{-\eta}$ is the path loss of the IRS-to-UT link with $d_r$ being the link distance. 
The LoS component in the IRS-to-UT channel is modeled as 
$\Bar{\mathbf{G}} = {\mathbf{a}}_\mathrm{IRS}{(\psi_a,\psi_e)}\,\mathbf{a}_\text{UT}({\psi_d})$, where $\psi_a$ and $\psi_e$ are the azimuth and elevation AoD of signals from the IRS, respectively, and ${\psi_d}$ is the AoA of the signals at the UT. The array response of 
$\mathbf{a}_\mathrm{IRS}{(\psi_a,\psi_e)}$ and $\mathbf{a}_\mathrm{UT}({\psi_d})$ are modeled as in (\ref{eq:a}) and (\ref{eq:b}), respectively.
\vspace{-1.65 em}
\subsection{Signal Model}\label{sec:1b}
Since BS exploits SSK and IRS exploits RPM to convey their local information to UT, the presented system can transmit a maximum of ${\rm{b}_{BS}} = \log_2(N_t)$ and ${\rm{b}_{IRS}} = \log_2(M)$ information bits from the BS and IRS, respectively. That is, the scheme can transmit  $\rm{b}=\rm{b}_{BS}+\rm{b}_{IRS}$ bits per channel use.
The received signal when the $t$-th TA is active at the BS with IRS reflecting the $m$-th phase shift is
\vspace{-0.5 em}
\begin{equation}\label{eq:d}
    \mathbf{y}=\sqrt{P_s}\mathbf{G}^H \mathbf{\Phi} \mathbf{H}\mathbf{x}+\mathbf{z}=\sqrt{P_s}{\sum_{n=1}^N}{h_{n,t}\mathbf{g}_n \text{exp}(j\varphi_m)}+\mathbf{z}\,,
\end{equation}
where $(\cdot)^H$ is the Hermitian, $\mathbf{y} \in \mathbb{C}^{N_r}$ is the received signal, $\mathbf{z} \in \mathbb{C}^{N_r}$ is the additive white Gaussian noise distributed as $\mathcal{CN}(0, \mathbf{I}_{N_r})$, $P_s$ is the total transmit power, $h_{n,t}$ is the $(n,t)$-th element of $\mathbf{H}$, and ${\mathbf{g}}_n$ denotes the Hermitian transpose of the $n$-\rm{th} row of $\mathbf{G}$. We have assumed that the components of $\mathbf{H}$, $\mathbf{G}$, and $\mathbf{z}$ are distributed independently. Furthermore, we utilize the ML detector to perform the joint detection of SSK and RPM symbols at the UT. The ML detector for (\ref{eq:d}) is 
\vspace{-0.25 em}
\begin{equation}\label{eq5}
    \{{\Hat{t},{{\varphi}_{\widehat{m}}}}\}= \arg \min_{{t},{\varphi_{m}}}\| \mathbf{y} -\sqrt{P_s}{\sum_{n=1}^N}{h_{n,t}\mathbf{g}_n \text{exp}(j\varphi_m)}\|^2\,,
\end{equation}
where ${\Hat{t}}\in[1,N_t]$ and ${\widehat{m}}\in[1,M]$. After determining the TA ${\Hat{t}}$ and reflection phase shift ${\varphi}_{\widehat{m}}$, the receiver can detect the information bits associated with the detected signal sets from Table \ref{tab:a}.
\vspace{-0.5 em}
\begin{table}[!t]
    \centering
    \caption{\textsc{Depicts the bit to constellation symbols mapping for $M=2$ and $N_t=2$}}
    \vspace{-0.5 em}
    \renewcommand{\arraystretch}{1}
    \setlength{\tabcolsep}{4pt}
    \begin{tabular}{|M{1cm}|M{0.5cm}|M{2.1cm}|M{1.7cm}|M{1.7cm}|}
        \hline
        {$\rm{b}_{BS}$} & {$\rm{b}_{IRS}$} & {Transmit vector $\mathbf{x}$} & {Phase shift $\varphi$} & {Set of signals} \\
       \hline 
       \vspace{0.25em} 
       0 & 0 & $\mathbf{i}_1=[1,0]^T$ & $\varphi_1$ & \{$\mathbf{i}_1,\varphi_1$\} \\
       \hline 
         \vspace{0.25em}
       1 & 0 & $\mathbf{i}_2=[0,1]^T$ & $\varphi_1$ & \{$\mathbf{i}_2,\varphi_1$\} \\
       \hline 
         \vspace{0.25em}
       0 & 1 & $\mathbf{i}_1=[1,0]^T$ & $\varphi_2$ & \{$\mathbf{i}_1,\varphi_2$\} \\
       \hline 
         \vspace{0.25em}
       1 & 1 & $\mathbf{i}_2=[0,1]^T$ & $\varphi_2$ & \{$\mathbf{i}_2,\varphi_2$\} \\
       \hline
    \end{tabular}
    \vspace{-0.15 em}
    \label{tab:a}
\end{table}

\section{Performance Analysis}
We derive the analytical framework of the ABER and ergodic capacity for the presented wireless network. The conditional PEP between the transmitted $\{t,\varphi_m\}$ and the detected signal sets $\{{\Hat{t}},{\varphi}_{\widehat{m}}\}$ can be given as
\vspace{-0.25 em}
\begin{eqnarray}\label{eq:e}
&&\mathrm{PEP}\left(\{t,\varphi_m\}\!\!\to\!\!\{{\Hat{t}},{\varphi}_{\widehat{m}}\}|\mathbf{G},\mathbf{H}\right) \nonumber\\ && =\Pr\left(\|\mathbf{y}-\sqrt{P_s}\lambda_{t,m}\|^2>\|\mathbf{y}-\sqrt{P_s}\lambda_{{\Hat{t}},{\widehat{m}}}\|^2\right)\,,
\end{eqnarray}
where $\lambda_{t,m}={\sum_{n=1}^N}{h_{n,t}\mathbf{g}_n\text{exp}(j\varphi_m)}$ and $\lambda_{\Hat{t},\widehat{m}}={\sum_{n=1}^N}{h_{n,\Hat{t}}\mathbf{g}_n\text{exp}(j{\varphi}_{\widehat{m}})}$. After solving (\ref{eq:e}), the conditional PEP is obtained as given by
\vspace{-0.5 em}
\begin{equation}\label{eq:f}
    \mathrm{PEP}\!\!\left(\!\{t,\varphi_m\}\!\!\to\!\!\{{\Hat{t}},{\varphi}_{\widehat{m}}\}\!\right)\!\!=\!\mathbb{E}\!\!\left[\!Q \left(\!\!\sqrt
    {\frac{P_s}{2}\|\lambda_{t,m}-\!\lambda_{{\Hat{t}},{\widehat{m}}}\|^2}\!\right)\!\right],
\end{equation}
where $\mathbb{E}[\cdot]$ is the expectation and $Q\left(\cdot\right)$ denotes the Gaussian $Q$-function. The average PEP can be determined by averaging 
(\ref{eq:f}) over Rician fading channels. The average PEP can then be utilized to derive the ABER of the presented system.
\subsection{ABER Analysis}
We derive the analytical expression for the ABER, which is a tight union bound \cite{singh_RIS_assisted_2022}, for the considered IRS-SSK-RPM scheme, given as
\begin{equation}\label{eq:pre_g}
{\mathrm{P_e}}\leq{\mathrm{P_e}}^{SSK}+{\mathrm{P_e}}^{RPM}+{\mathrm{P_e}}^{SSK-RPM}\,, 
\end{equation}
where
\vspace{-0.5 em}
\begin{eqnarray}\label{eq:g}
&&\mathrm{P_e}^{SSK}= \frac{\sum_{t,\Hat{t}=1}^{N_t} \mathcal{D}_H(t,\Hat{t})
\mathrm{PEP}(\{t\!\!\to\!\!\Hat{t} \})}{N_t\log_2(N_tM)}\,,\nonumber\\
&&\mathrm{P_e}^{RPM} = \frac{\sum_{m,\widehat{m}=1}^M \mathcal{D}_H(\varphi_{m},{\varphi}_{\widehat{m}})\mathrm{PEP}(\{\varphi_m \!\!\to\!\!{\varphi}_{\widehat{m}}\})}{M\log_2(N_tM)} \,,\nonumber\\ 
&&\mathrm{P_e}^{SSK-RPM}= \frac{1}{MN_t\log_2(N_tM)}\sum_{m,\widehat{m} \ne m=1}^M \sum_{{t,\Hat{t}\ne t}=1}^{N_t}\nonumber\\ 
&& \left[\mathcal{D}_H(t,\Hat{t})\!\!+\!\mathcal{D}_H(\varphi_{m},{\varphi}_{\widehat{m}})\right]\!\mathrm{PEP}(\{t,\varphi_m \}\!\!\to\!\!\{ \Hat{t},{\varphi}_{\widehat{m}}\}),
\end{eqnarray}
where $\mathcal{D}_H(\cdot,\cdot)$ denotes the hamming distance between different symbols. $\mathrm{P_e}^{SSK}$ represents the error probability associated with the antenna index when the reflection phase shift is detected correctly ($m \rightarrow {\widehat{m}}$). $\mathrm{P_e}^{RPM}$ denotes the error probability for RPM constellation symbols when the active TA index is detected correctly $(t\rightarrow\Hat{t})$. $\mathrm{P_e}^{SSK-RPM}$ indicates the error probability when the antenna index and the RPM constellation symbols are detected incorrectly.

To determine $\mathrm{P_e}^{SSK}$, we need to derive the average $\mathrm{PEP}\left(\{t \to \Hat{t}\}\right)$ between the transmit and receive antenna indices when the IRS reflects the incoming signal with the $m$-th RPM constellation symbol as given by 
\begin{equation}\label{eq:h}
\mathrm{PEP}\left(\!\{t\!\!\rightarrow\!\!\Hat{t}\}\!\right) \!\! = \!\! \frac{1}{M} \sum_{m=1}^{M}\int_0^{\infty} {Q\left(\sqrt{\frac{P_s}{2} \, \xi_m}\right)}{f(\xi_m)}\text{d}{\xi_m}\,,
\end{equation}
where ${\xi_m}=\|\lambda_{t,m}-\lambda_{{\Hat{t}},{{m}}}\|^2= \|{\sum_{n=1}^N}{h_{n,t}\mathbf{g}_n\text{e}^{j\varphi_m}}-{\sum_{n=1}^N}{h_{n,{\Hat{t}}}\mathbf{g}_n\text{e}^{j\varphi_m}}\|^2$. By expanding $\mathbf{g}_n$, where $g_{n,r}$ is the $r$-th element of $\mathbf{g}_n$ and $r\in[1,N_r]$, we obtain
\begin{equation}\label{eq:i}
{\xi_m}=\|\mathbf{u}\|^2=|u_1|^2+|u_2|^2+\dots+| u_{N_r}|^2\,,
\end{equation}
where $u_r={\sum_{n=1}^N}{h_{n,t}g_{n,r}\text{e}^{j\varphi_m}}-{\sum_{n=1}^N}{h_{n,{\Hat{t}}}g_{n,r}\text{e}^{j\varphi_m}}$. The $(n,r)$-th element of  
$g_{n,r} = \Re{(g_{n,r})}+j\Im(g_{n,r})$,  where $\Re(\cdot)$ and $\Im (\cdot)$ denotes the real and imaginary components, respectively. Further, $\left| u_r \right|^2 = |\alpha_{r}|^2 + |\beta_{r}|^2$, where $\alpha_r=\sum_{n=1}^N \left(\Re(g_{n,r})  \left( h_{n,t}-h_{n,\hat{t}} \right)\right)$ and $\beta_r=\sum_{n=1}^N\left(\Im(g_{n,r})\left( h_{n,t}-h_{n,\hat{t}} \right)\right)$. 
From (\ref{eq:c}), the mean of $g_{n,r}$  is given by $\mathbb{E}[\Re(g_{n,r})]=\sqrt{\frac{K_r\nu_r}{1+K_r}}{{\bar{g}_{n,r}}}$, where $\bar{g}_{n,r}$ is the $(n,r)$-th element of $\bar{{\mathbf{G}}}$ and $\mathbb{E}[\Im(g_{n,r})]=0$. The variance of $g_{n,r}$ from (\ref{eq:c}) can be obtained as $\mathbb{V}[\Re(g_{n,r})]= {\frac{\nu_r}{2(1+K_r)}}$ and  $\mathbb{V}[\Im(g_{n,r})]={\frac{\nu_r}{2(1+K_r)}}$, where $\mathbb{V}[\cdot]$ denotes the variance. Thus, $\alpha_r\sim{\cal{N}}(\mathbb{E}[\alpha_r], \sigma_m^2)$ and $\beta_r\sim {\cal{N}}(\mathbb{E}[\beta_r],\sigma_m^2)$ are Gaussian distributed RVs, where mean is $\mathbb{E}[\alpha_r]=\sqrt{\frac{K_r\nu_r}{1+K_r}} \sum_{n=1}^N \bar{g}_{n,r}\left( h_{n,t}-h_{n,\hat{t}} \right)$, $\mathbb{E}[\beta_r]=0$, and variance is $\sigma_m^2 = {\frac{\nu_r}{2(1+K_r)}} \sum_{n=1}^N \left|h_{n,t}-h_{n,\hat{t}} \right|^2$. 
If $|\alpha_{r}+j \beta_r|$ follows Rician distribution, then from (\ref{eq:i}), $\xi_m$ represents the sum of the squares of $N_r$ Rician distributed RVs. This can be modeled as a non-central chi-square distribution with $2N_r$ degrees of freedom (DoF) and the probability density function (PDF) \cite{hu_accurate_close_2005} can be written as 
\begin{equation}\label{eq:j}
f(\xi_m)\!=\!\frac{1}{2\sigma_m^2}{\left( \frac{\xi_m}{s_{m}^2}\right)}^{\frac{N_r-1}{2}}\!{\text{e}}^{{-\frac{\xi_m+s_{m}^2}{2\sigma_m^2}}}I_{\small{N_r-1}}\!\!\left(\frac{\sqrt{\xi_m}s_{m}}{\sigma_m^2}\right)\,,
\end{equation}
where $\xi_m > 0$, $I_{N_r-1}(\cdot)$ denotes the modified bessel function of $(N_r-1)$-th order, and $s_{m}=\sqrt{\sum_{r=1}^{N_r}|\mathbb{E}[\alpha_r]|^2}$.  
After substituting (\ref{eq:j}) in (\ref{eq:h}) and by using an alternating form of $Q$-function, that is Craig's formula, and by applying the MGF-based approach \cite{Yigit_Hybrid_Reflection_2023}, the 
average PEP can be written as 
\begin{equation}\label{eq:k}
   \mathrm{PEP}\left(\! \{t \!\! \to \!\! \Hat{t} \}\! \right)\! =\! \frac{1}{M} \sum_{m=1}^{M} \frac{1}{\pi}\int_0^{\frac{\pi}{2}} \mathcal{M}_{\xi_m}
    \left( \frac{P_s \xi_m}{2\sin^2{\Omega}}\right)\text{d}\Omega \,,
\end{equation}
where $\mathcal{M}_{\xi_m}\left( \frac{P_s \xi_m}{2\sin^2{\Omega}}\right)=\int_0^{\infty}\exp{\left( -\frac{P_s \xi_m}{2{\sin^2{\Omega}}}\right)\!f(\xi_m)\,\text{d}\xi_m}$ is the MGF. 
The MGF of $\xi_m$ with $2N_r$ DoF \cite{Proakis2007} can be obtained as  
\vspace{-0.5 em}
\begin{equation}\label{eq:l}
    \mathcal{M}_{\xi_m}(\omega)=\left(\frac{1}{1-2\omega\sigma_m^2}\right)^{Nr} \exp\left(\frac{\omega s_m^2}{1-2\omega \sigma_m^2}\right)\,.
\end{equation}
By substituting (\ref{eq:l}) in (\ref{eq:k}) and by solving the integral using \cite{chiani_New_exponential_2003}, 
we can approximate the average PEP in (\ref{eq:k}) as 
\begin{equation}\label{eq:k1}
    \mathrm{PEP}\left(\! \{t \!\! \to \!\! \Hat{t} \}\! \right) \!\! \approx \!\! \frac{1}{12}\mathcal{M}_{\xi_m}\left( \frac{P_s \xi_m}{2}\right)\!+\!\frac{1}{4}\mathcal{M}_{\xi_m}\left( \frac{2P_s \xi_m}{3}\right)\,.
\end{equation}
Thus, \eqref{eq:k1} is utilized in (\ref{eq:g}) to determine $\mathrm{P_e}^{SSK}$.

The average PEP between the RPM constellation symbols when the TA index is known to the receiver is given by
\begin{equation}\label{eq:m}
    \mathrm{PEP}\left(\!\{\varphi_m \!\!\to\!\! {\varphi}_{\widehat{m}}\} \!\right) \!\!=\!\! \frac{1}{N_t}\!\! \sum_{t=1}^{N_t} \int_0^{\infty} Q\left(\sqrt{\frac{P_s}{2}{\xi}_t}\right)\!\!f({\xi}_t)\,\text{d}{\xi}_t\,,
\end{equation}
where ${\xi}_t= \|\lambda_{t,m}-\lambda_{{{t}},{\widehat{m}}}\|^2$. Similar to (\ref{eq:i}), ${\xi}_t = |v_1|^2+|v_2|^2+\dots+| v_{N_r}|^2$, where $v_r = {\sum_{n=1}^N}{h_{n,t}{g}_{n,r}\text{e}^{j\varphi_m}}-{\sum_{n=1}^N}{h_{n,{{t}}}{g}_{n,r}\text{e}^{j{\varphi}_{\widehat{m}}}}$, and $\left| v_r \right|^2 = |a_{r}|^2 + |b_{r}|^2$. 
%with $a_{r}=\sum_{n=1}^N {h_{n,t}(\Re(g_{n,r})(\cos\varphi_m-\cos{{\varphi}_{\widehat{m}}})}-\Im(g_{n,r})(\sin\varphi_m-\sin{\varphi_{\widehat{m}}}))$ and $b_{r}=\sum_{n=1}^N {h_{n,t}(\Im(g_{n,r})(\cos\varphi_m-\cos{{\varphi}_{\widehat{m}}})}-\Re(g_{n,r})(\sin\varphi_m-\sin{\varphi_{\widehat{m}}}))$. 
The RVs $a_r\sim{\mathcal{N}}(\mathbb{E}[a_r], \sigma_t^2)$ and $b_r\sim{\mathcal{N}}(\mathbb{E}[b_r],\sigma_t^2)$ are Gaussian distributed with $\mathbb{E}[a_r]=\sqrt{\frac{K_r\nu_r}{1+K_r}}(\cos\varphi_m-\cos{{\varphi}_{\widehat{m}}})\sum_{n=1}^N{\Bar{g}_{n,r}h_{n,t}}$, $\mathbb{E}[b_r]=\sqrt{\frac{K_r\nu_r}{1+K_r}}(\sin\varphi_m-\sin{{\varphi}_{\widehat{m}}})\sum_{n=1}^N{\Bar{g}_{n,r}h_{n,t}}$, and $\sigma_t^2= {\frac{\nu_r}{(1+K_r)}(1-\cos(\varphi_m-{\varphi}_{\widehat{m}}))\sum_{n=1}^N \left|h_{n,t}\right|^2}$. The RV ${\xi}_t$ can be modeled as non-central chi-square distribution with $2N_r$ DoF with non-centrality parameter $s_t=\sqrt{\sum_{r=1}^{N_r}\left(|\mathbb{E}[a_r]|^2 +|\mathbb{E}[b_r]|^2\right)}$ and variance $\sigma_t^2$.  
By following the MGF-based approach as given before in (\ref{eq:k}) and then substituting the MGF of $\xi_t$ using (\ref{eq:l}) in (\ref{eq:m}) the $\mathrm{PEP}(\varphi_m \!\!\to\!\! {\varphi}_{\widehat{m}})$ can be obtained. This is further used in (\ref{eq:g}) to determine $\mathrm{P_e}^{RPM}$. 

The joint $\mathrm{PEP}\left(\!\{t,\varphi_m\}\!\! \to \!\! \{\Hat{t},{\varphi}_{\widehat{m}}\}\!\right)$ can be expressed as
\vspace{-0.5 em}
\begin{equation}\label{eq:q}
    \mathrm{PEP}\left(\!\{t,\varphi_m\}\!\! \to \!\! \{\Hat{t},{\varphi}_{\widehat{m}}\}\!\right)\!\!=\!\! \int_0^{\infty} Q\left(\sqrt{\frac{P_s}{2}{\xi}}\right) f({\xi})\text{d}{\xi}\,,
\end{equation}
where ${\xi}= \|\lambda_{t,m}-\lambda_{{\Hat{t}},{\widehat{m}}}\|^2$. 
Similar to (\ref{eq:i}), ${\xi}=|w_1|^2+|w_2|^2+\dots+| w_{N_r}|^2$, where  $w_r={\sum_{n=1}^N}{h_{n,t}{g}_{n,r}\text{exp}(j\varphi_m)}-{\sum_{n=1}^N}{h_{n,{\Hat{t}}}{g}_{n,r}\text{exp}(j{\varphi}_{\widehat{m}})}$ and $|w_r|^2=|c_{r}|^2 + |e_{r}|^2$. 
%edited ****
%\vspace*{-0.35em}
%\begin{equation*}
%\begin{split}
%c_{r}=\left(\sum_{n=1}^N {({h_{n,t}-h_{n,\Hat{t}})\Re(g_{n,r})(\cos\varphi_m-\cos{{\varphi}_{\widehat{m}}})}}\right.\\\left.-\sum_{n=1}^N (h_{n,t}-h_{n,\Hat{t}})\Im(g_{n,r})(\sin\varphi_m-\sin{{\varphi}_{\widehat{m}}})\right)\\ e_{r}=\left(\sum_{n=1}^N {({h_{n,t}-h_{n,\Hat{t}})\Im(g_{n,r})(\cos\varphi_m-\cos{{\varphi}_{\widehat{m}}})}}\right. \\\left.+\sum_{n=1}^N(h_{n,t}-h_{n,\Hat{t}})\Re(g_{n,r})(\sin\varphi_m-\sin{{\varphi}_{\widehat{m}}})\right) \, .
%\end{split}
%\end{equation*}
The RVs $c_r \!\!\!\!\sim\!\!\!\!{\mathcal{N}}(\mathbb{E}[c_r], \sigma^2)$ and $e_r \!\!\sim\!\!{\mathcal{N}}(\mathbb{E}[e_r],\sigma^2)$ are Gaussian distributed with $\mathbb{E}[c_{r}]= \sqrt{\frac{K_r\nu_r}{1+K_r}}(\cos\varphi_m-\cos{{\varphi}_{\widehat{m}}})\sum_{n=1}^N{\left(\Bar{g}_{n,r}(h_{n,t}-h_{n,\Hat{t}})\right)},\mathbb{E}[e_r]=~~\sqrt{\frac{K_r\nu_r}{1+K_r}}(\sin\varphi_m-\sin{{\varphi}_{\widehat{m}}})\sum_{n=1}^N{\left(\Bar{g}_{n,r}(h_{n,t}-h_{n,\Hat{t}})\right)}$, and $\sigma^2 =~~\frac{\nu_r}{(1+K_r)} (1- \cos(\varphi_m-{\varphi}_{\widehat{m}}))\sum_{n=1}^N \left(\left|h_{n,t}-h_{n,\Hat{t}}\right|^2\right)$. The RV ${\xi}$ follows non-central chi-square distribution with $2N_r$ DoF \cite{hu_accurate_close_2005}, where the non-centrality parameter $s=\sqrt{\sum_{r=1}^{N_r} \left(|\mathbb{E}[c_{r}]|^2 + |\mathbb{E}[e_{r}]|^2\right)}$ and variance $\sigma^2$.
As before, by using the MGF-based approach given in (\ref{eq:k}) and then substituting the MGF of $\xi$ using (\ref{eq:l}) in (\ref{eq:q}), we obtain the closed-form expression for the average $\mathrm{PEP}(\{t,\varphi_m\} \to \{\Hat{t},{\varphi}_{\widehat{m}}\})$. This is substituted in (\ref{eq:g}) to determine $\mathrm{P_e}^{SSK-RPM}$.
Therefore, the derived average PEPs, i.e., (\ref{eq:h}), (\ref{eq:m}), and (\ref{eq:q}) are employed in (\ref{eq:g}) to determine the final expression of ABER as given by (\ref{eq:pre_g}).
\vspace{-1.25 em}
\subsection{Diversity Analysis}
The values of $\mathrm{P_e}^{SSK}$ and $\mathrm{P_e}^{RPM}$  are less than the $\mathrm{P_e}^{SSK-RPM}$ as can be observed from (\ref{eq:g}). Therefore, the ABER of the considered system is similar to $\mathrm{P_e}^{SSK-RPM}$ in the high signal-to-noise ratio (SNR) regime. From (\ref{eq:g}), the  $\mathrm{P_e}^{SSK-RPM}$ is directly related to the PEP; we obtain the asymptotic behaviour at high SNR as given by 
\vspace{-0.25 em}
\begin{equation}\label{eq:t}
    \mathrm{PEP}(\{t,\varphi_m\} \to \{\Hat{t},{\varphi}_{\widehat{m}}\}) \propto \left( \frac{1}{P_s\sigma^2}\right)^{N_r}\,.
\end{equation}
Thus, from (\ref{eq:t}), the diversity order is $N_r$.
\vspace{-1.25 em}
\subsection{Ergodic Capacity Analysis}
\vspace{-0.5em}
The discrete input and continuous output memoryless channel (DCMC) capacity is given by \cite{singh_RIS_assisted_2022}
\begin{equation}\label{eq:u}
    \mathrm{C}=\mathbb{E}\left[ {\max_{f(\mathbf{x}),f(\varphi)}} \,\mathcal{I} (\mathbf{x},\varphi;\mathbf{y})\right]\,,
\end{equation}
where $\mathcal{I} (\mathbf{x},\varphi;\mathbf{y})$ represents the mutual information between the signal set $\{{\mathbf{x},\varphi}\}$ and the received vector $\mathbf{y}$. By applying the chain rule, we can express $  \mathcal{I} (\mathbf{x},\varphi;\mathbf{y})$ as
\vspace*{-0.25em}
\begin{equation}\label{eq:v}
    \mathcal{I} (\mathbf{x},\varphi;\mathbf{y})= \mathcal{I}(\varphi;\mathbf{y})+ \mathcal{I}(\mathbf{x};\mathbf{y}|\varphi)\,,
   % \vspace*{-0.25em}
\end{equation}
where $\mathcal{I}(\varphi;\mathbf{y})$ is the mutual information between $\mathbf{y}$ and $\varphi$, and $\mathcal{I}(\mathbf{x};\mathbf{y}|\varphi)$ represents the information that $\mathbf{y}$ carries regarding $\mathbf{x}$ given $\varphi$. $\mathcal{I}(\varphi;\mathbf{y})$ is defined as
\vspace*{-0.45em}
\begin{eqnarray}\label{eq:w}
\mathcal{I}(\varphi;\mathbf{y})=\underbrace {\sum_{\varphi}\int_\mathbf{y}  f(\varphi)f(\mathbf{y}|\varphi)\log_2 f(\mathbf{y}|\varphi)\text{d}\mathbf{y}}_{\gamma_1}\nonumber\\
-\underbrace{\sum_{\varphi}\int_\mathbf{y} f(\varphi)f(\mathbf{y}|\varphi)\log_2 f(\mathbf{y})\text{d}\mathbf{y}}_{\gamma_2}\,,
\end{eqnarray}
where $f(\cdot|\cdot)$ represents the conditional PDF. $\mathcal{I}(\mathbf{x};\mathbf{y}|\varphi)$ is expressed as
\vspace*{-0.65em}
\begin{eqnarray}\label{eq:x}
\mathcal{I}(\mathbf{x};\mathbf{y}|\varphi)= \underbrace{\sum_{\varphi}\sum_{\mathbf{x}}\int_{\mathbf{y}} f(\mathbf{x},\mathbf{y}|\varphi)\log_2f(\mathbf{y}|\mathbf{x},\varphi)\text{d}\mathbf{y}}_{\Gamma_1}\nonumber\\-\underbrace{\sum_{\varphi}\sum_{\mathbf{x}}\int_{\mathbf{y}} f(\mathbf{x},\mathbf{y}|\varphi)\log_2f(\mathbf{y}|\varphi)\text{d}\mathbf{y}}_{\Gamma_2}\,.
\end{eqnarray}
After substituting the conditional PDFs in (\ref{eq:w}) and (\ref{eq:x}), it becomes apparent that $\gamma_1$ and  $\Gamma_2$ are equal. Consequently, by substituting the values from (\ref{eq:w}) and (\ref{eq:x}) into (\ref{eq:v}), $\mathcal{I}(\mathbf{x},\varphi;\mathbf{y})$ for the considered system is obtained as
\begin{equation}\label{eq:y}
\begin{split}
\mathcal{I}(\mathbf{x},\varphi;\mathbf{y})&\!\approx\! 2\log_2(N_tM)\\
&-\!\log_2 \left(\!\!N_tM \!+\! \!\!\sum_{m,\widehat{m}=1}^M\! \sum_{\Hat{t}, t=1}^{N_t} \!\mathbb{E}\!\! \left[ \exp\left(-\frac{P_s\xi}{2}\! \right)\! \right] \!\right)\!,
\end{split}
\end{equation}
where $\widehat{m}\ne m$, $\Hat{t}\ne t$, and $\xi=\| \lambda_{t,m} - \lambda_{\Hat{t},\widehat{m}} \|^2$.
The $\mathbb{E} \left[ \exp{(-\frac{P_s}{2}\xi)}\right]$ can be written as the MGF of $\xi$ and by substituting (\ref{eq:y}) in (\ref{eq:u}), we obtain the expression for $\mathrm{C}$ as
\begin{eqnarray}\label{eq:z}
\mathrm{C}=2\log_2(N_tM)-\log_2\left[ N_tM + \sum_{m=1}^M \sum_{m \ne {\widehat{m}}=1}^M\right.\nonumber \\ \left.   \sum_{t=1}^{N_t} \sum_{t \ne {\Hat{t}} =1}^{N_t} \mathcal{M}_\xi\left({\frac{P_s}{2} \xi}\right) \right]\,.
\end{eqnarray}
\section{Results and discussion}
\begin{figure}[!t]
    \centering
    \includegraphics[width=3.10 in, height=2.2 in]{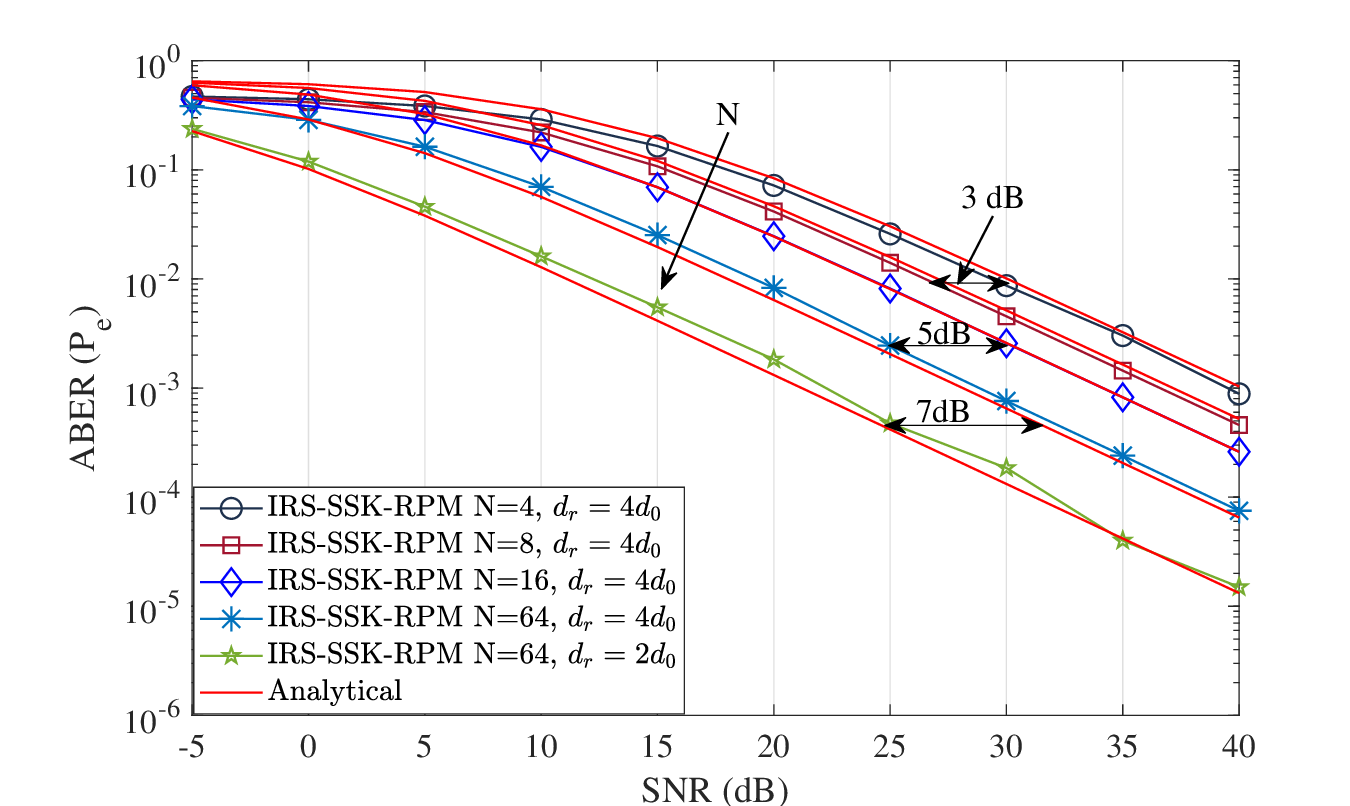}
    \caption{\small ABER vs SNRs for $N_t=2$, $N_r=1$, $M=2$, and various values of $N$.
    }
    \label{fig: figure2}
\end{figure}

We discuss numerical results for the derived analytical results of the IRS-SSK-RPM scheme over Rician channels. 
In the results, we assume $d_0=1$ km, $d_t=d_0$, $\eta=2.3$, $\rho_0=1$, and $K_r=2$. All the results have been verified extensively through Monte-Carlo simulations.  Fig. \ref{fig: figure2} depicts the performance of the ABER versus SNR for the proposed system for different values of reflecting elements $N$. We observe from Fig. \ref{fig: figure2} that the analytical and simulation results of ABER closely follow in higher SNR regimes, whereas in lower SNR regions, they establish a tight bound. The performance of ABER of the presented system improves with increasing $N$. For a fixed value of $\mathrm{P}_e=10^{-2}$, when $N$ is doubled, approximately $3\,\mathrm{dB}$ array gain is achievable. Furthermore, we observe a $7\,\mathrm{dB}$ array gain when the $d_r$ is reduced by half at $\mathrm{P}_e \approx10^{-3}$. This observation emphasizes the superiority of utilizing IRS in the SSK modulation over the Rician channel to improve the ABER performance. 

\begin{figure}[!t]
    \centering
    \includegraphics[width=3.10 in, height=2.2 in]{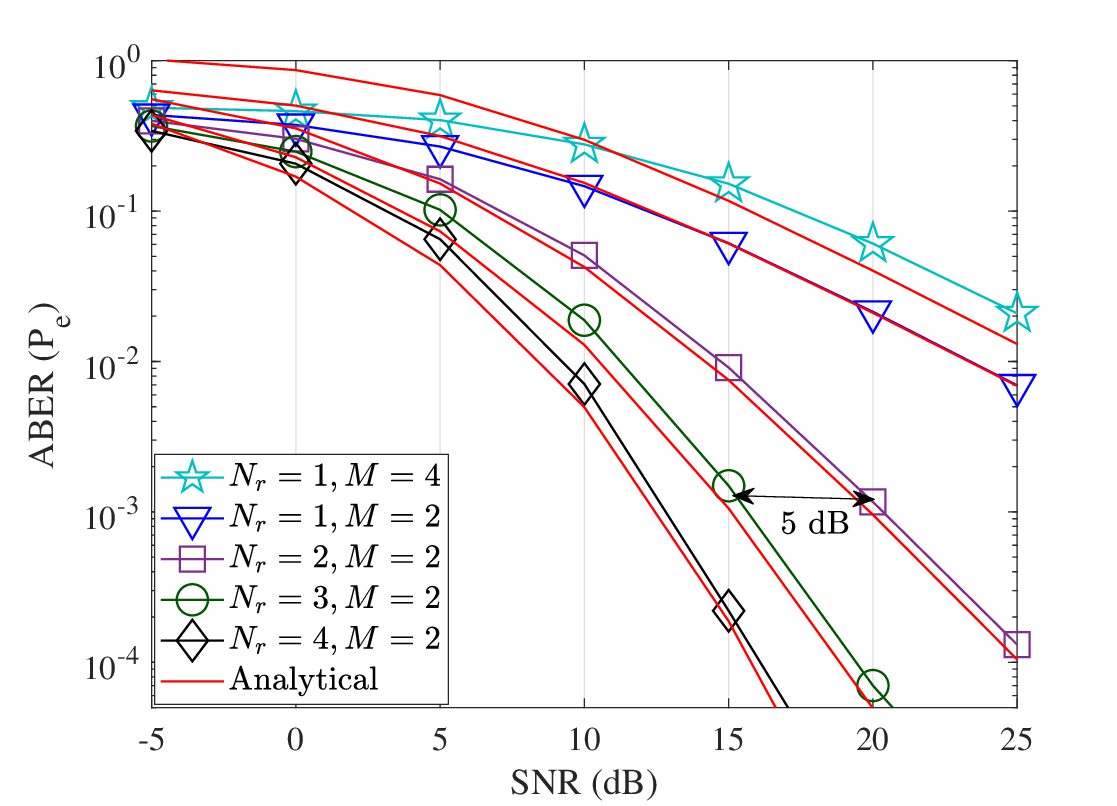}
    \caption{\small  ABER as a function of SNRs  for $N_t=2$, $N=20$, $d_r=4d_0$ and increasing values of $N_r$ and $M$. 
    }
    \label{fig: figure3}
\end{figure}

We further illustrate the ABER versus SNR plots in Fig. \ref{fig: figure3} for  $N=20$, $N_t=2$, $d_r=4d_0$ and different numbers of $N_r$.  We can observe from the figure that for ABER of $10^{-3}$ with an increase in $N_r$ from $2$ to $3$, the system attains $5\,\mathrm{dB}$ SNR gain. For $N_r = M = 2$, it can be observed from ABER plots that the values of ABER are $10^{-2}$ and $10^{-4}$ at 15 dB and 25 dB SNRs, respectively. Therefore, the diversity order of the presented system is equal to $\text{log}(10^{-2})-\text{log}(10^{-4})=2$, which validates the analytically derived diversity order. A slight difference is observed between the analytical results and simulations due to the assumptions considered to obtain the ABER expression. Our results demonstrate that we can substantially enhance the performance by adjusting the system parameters, such as the $N$ and increasing the number of $N_r$ at the UT.

\begin{figure}[!t]
    \centering
    \includegraphics[width=3.10 in, height=2.2 in]{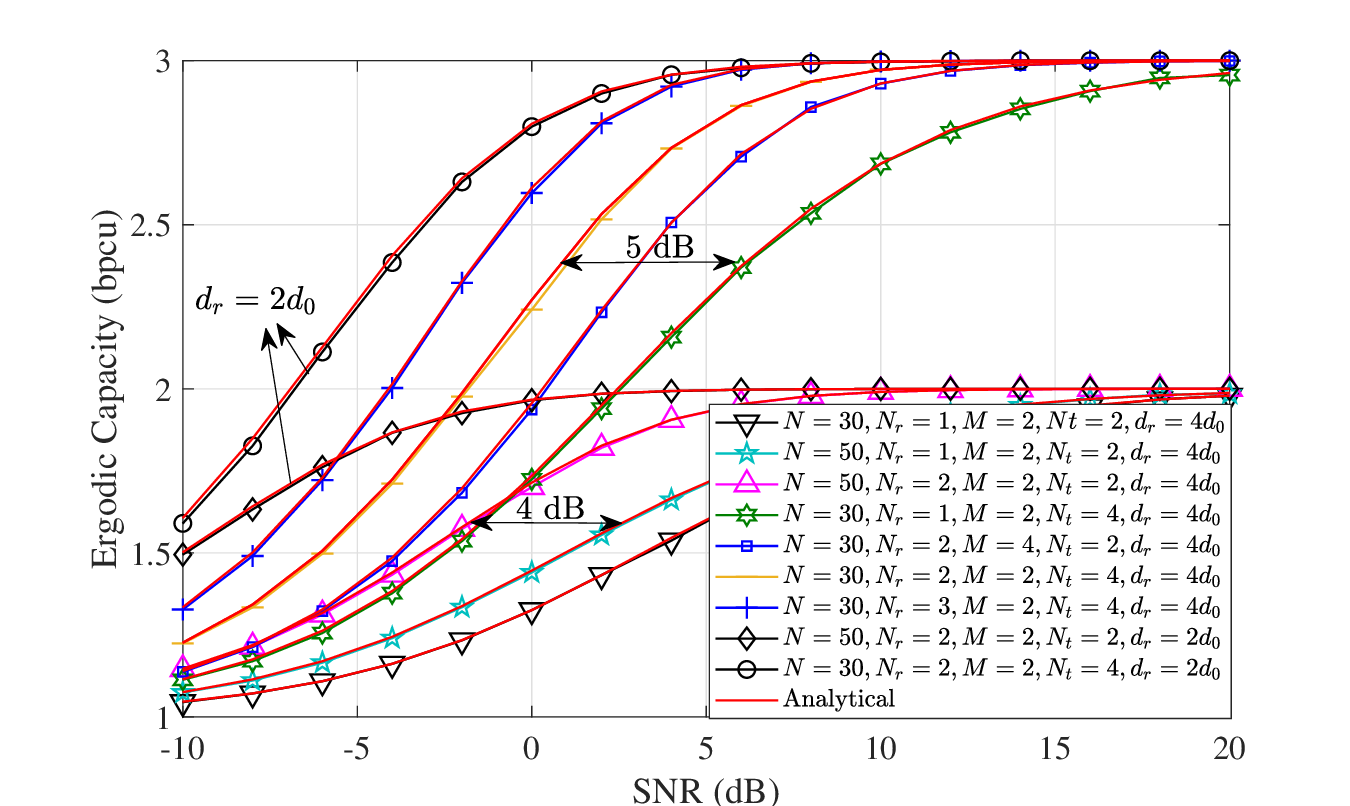}
    \caption{\small Ergodic capacity vs SNRs for various values of $M$, $N_t$, $N_r$, and $N$.
    }
    \label{fig: figure4}
\end{figure}

Ergodic capacity versus SNR is illustrated in Fig. \ref{fig: figure4} for different sizes of SSK and RPM constellations. In the low SNR regime, ergodic capacity exhibits logarithmic growth, and in the high SNR regime, it converges towards the upper limit of \, $\log_2(N_tM)$. We observe from the figure that for fixed SE, the system achieves array gain with increasing $N$. It is also noticed from the figure that larger values of $N_r$ lead to reaching the upper limit at lower SNR. Specifically, we can see a $4\,\, \mathrm{dB}$ SNR gain when $N_r$ changes from $1$ to $2$ for fixed $N$ and SE. These results emphasize that in the low SNR regime, the ergodic capacity of the IRS-SSK-RPM scheme exhibits improvement with increasing values of $N_r$ and $N$. Noticing that increasing $K_r$ strengthens the LoS component in (\ref{eq:c}) enhances the SNR but decreases spatial diversity. As a result, we observed a negative impact on ABER and ergodic capacity by increasing $K_r$. On the other hand, a decrease in $d_r$ enhances the SNR while maintaining spatial diversity, which improves both ABER and ergodic capacity. 

\section{Conclusion}
We have thoroughly analyzed the performance of the IRS-SSK-RPM scheme over Rician fading channels. Through our presented system, we have shown the utilization of SSK modulation and reflection phase modulation using IRS, demonstrating the combined advantages of the system's performance. Maximum-likelihood detection has been employed to jointly detect the transmitting antenna index and the reflection phase shift symbol, and further, a closed-form expression for $\mathrm{PEP}$ has been derived for the Rician fading. In particular, we have derived the closed-form analytical framework for the considered system in terms of ABER and achievable ergodic capacity. The IRS-SSK-RPM scheme significantly enhances system reliability and ergodic capacity by enabling the direct transmission of information through reflection phase shifts while concurrently reflecting incoming wireless signals. The derived metrics, such as the ABER and the ergodic capacity, are utilized to investigate the system performance in detail as a function of the transmitting antennas, reflecting elements, receive antennas, and the number of constellation symbols.

\bibliographystyle{IEEEtran}
\bibliography{references}
\end{document}